\documentclass[twocolumn,english,aps,prl,reprint,superscriptaddress]{revtex4}
\usepackage[LGR,T1]{fontenc}
\usepackage[latin9]{inputenc}
\usepackage{verbatim}
\usepackage{amstext}
\usepackage{amssymb}
\usepackage{graphicx}
\usepackage{xcolor}

\makeatletter

\ProvideTextCommand{\~}{LGR}[1]{\char126#1}

\usepackage[english]{babel}
\def\urlprefix{}
\def\url#1{}
\addto\captionsenglish{%
}

\usepackage[none]{hyphenat}

\makeatother

\def\T{{\, \textrm{T}}}

\begin{document}
\title{Three Jahn-Teller states of matter in the spin-crossover system Mn(taa)}
\author{Jie-Xiang Yu}

\affiliation{Department of Physics, Center for Molecular Magnetic Quantum Materials
and Quantum Theory Project, University of Florida, Gainesville, Florida
32611, USA}
\author{Dian-Teng Chen}
\affiliation{Department of Physics, Center for Molecular Magnetic Quantum Materials
and Quantum Theory Project, University of Florida, Gainesville, Florida
32611, USA}
\author{Jie Gu}
\affiliation{Department of Physics, Center for Molecular Magnetic Quantum Materials
and Quantum Theory Project, University of Florida, Gainesville, Florida
32611, USA}
\author{Jia Chen}
\affiliation{Department of Physics, Center for Molecular Magnetic Quantum Materials
and Quantum Theory Project, University of Florida, Gainesville, Florida
32611, USA}
\author{Jun Jiang}
\affiliation{Department of Physics, Center for Molecular Magnetic Quantum Materials
and Quantum Theory Project, University of Florida, Gainesville, Florida
32611, USA}
\author{Long Zhang}
\affiliation{Department of Physics, Center for Molecular Magnetic Quantum Materials
and Quantum Theory Project, University of Florida, Gainesville, Florida
32611, USA}
\author{Yue Yu}
\affiliation{Department of Physics, Center for Molecular Magnetic Quantum Materials
and Quantum Theory Project, University of Florida, Gainesville, Florida
32611, USA}
\author{Xiao-Guang Zhang}
\affiliation{Department of Physics, Center for Molecular Magnetic Quantum Materials
and Quantum Theory Project, University of Florida, Gainesville, Florida
32611, USA}
\author{Vivien S. Zapf}
\affiliation{National High Magnetic Field Lab (NHMFL), Los Alamos National Lab
(LANL), Los Alamos NM 87545, USA}
\author{Hai-Ping Cheng}
\thanks{Correspond to: hping@ufl.edu}

\affiliation{Department of Physics, Center for Molecular Magnetic Quantum Materials
and Quantum Theory Project, University of Florida, Gainesville, Florida
32611, USA}

\begin{abstract}
Three high-spin phases recently discovered in the spin-crossover system Mn(taa) 
are identified through analysis by a combination of first-principles calculations 
and Monte Carlo simulation as a low-temperature Jahn-Teller ordered (solid) phase, an intermediate-temperature 
dynamically correlated (liquid) phase, and an uncorrelated (gas) phase.
In particular, the Jahn-Teller liquid phase arises from competition between mixing with low-spin impurities, 
which drive the disorder, and inter-molecular strain interactions.
The latter are a key factor in both the spin-crossover phase transition and the magnetoelectric coupling.
Jahn-Teller liquids may exist in other spin-crossover materials and materials 
that have multiple equivalent Jahn-Teller axes.
\end{abstract}
\maketitle


Jahn Teller (JT) distortions play important roles in many spin-crossovers (SCO) (also known as spin state transitions) in molecule-based magnets \cite{Decurtins_1984a,Sim_1981a,Halcrow_2011a,Matsumoto_2014a,Fitzpatrick_2015a} 
In SCO, electrons in partially filled $d$ shells transition between orbitals to change the overall spin state.
Long-range ordering of the JT distortions in the S = 2 HS state of crystalline $[\mathrm{Mn}^{3+}(\mathrm{pyrol})_{3}(\mathrm{tren})]$
(Mn(taa)) has been shown \cite{Chikara_2019a} to lead to magnetoelectric coupling, 
which is the interplay between magnetism and electric polarization or dielectric properties. 
Magnetoelectric coupling offers potential for applications in low-power magnetic sensors, 
frequency devices, data storage and other applications for
moving beyond Moore's law.\cite{Bea_2008,Fusil_2014a,Fiebig_2016a}
Molecule-based magnetic materials have
recently stimulated much interest because of the global initiative
in quantum information science and spintronics, \cite{Leuenberger_2001a,Bogani_2008a}
and their low Young's modulus offers great potential for lattice-mediated magnetoelectric coupling.

In conventional solids such as inorganic perovskites, cooperative JT effects are linked to structural and metal-insulator phase transitions \cite{Gehring_1975,Kaplan}, and are crucial in explaining many physical properties of materials such as colossal magnetoresistance \cite{PhysRevB.88.174426}. 
In the molecular solid Mn(taa), JT distortions are also expected to show long-range correlation due to the inter-molecular strain field. 
In the high spin (HS) state,  three nearly-degenerate JT distortions along axes $120^{\circ}$ apart reduce the Mn site symmetry from $C_{3}$ to $C_{1}$, leading to electric dipoles not cancelled by symmetry. 
In Ref. \cite{Chikara_2019a}, a modified four-state Ising-Potts model\cite{Nakano_2003a,Stratt_1987a}
based on a mean-field approach\cite{Boukheddaden_2000a,Boukheddaden_2000b} was used to explain the magnetoelectric behavior in terms of ordering of the three JT distortions and their associated electric dipoles. 
However, it did not offer a microscopic model for the inter-molecular interactions, nor an exact long-range ordering pattern of JT distortions. 
Rather, it described the emergence of three HS phases in terms of the statistical distributions of three degenerate JT distortion axes. 

In this Letter, we use a theoretical approach that combines first-principles calculations and Monte-Carlo simulations to study the microscopic
process underlying the phase transition and magnetoelectric coupling of Mn(taa), 
explicitly including the inter-molecular correlation of the JT distortion. 
Through analysis of the correlation function and polarization fluctuation, 
we show that the three HS phases discovered experimentally correspond to a solid-like phase, where the JT distortions are ordered and uniform, a liquid-like phase, where the JT distortions are disordered with long range correlation, and a gas-like phase, where the JT distortions are uncorrelated. 
In particular, the JT liquid phase is driven by a competition between disorder, which is caused in part by dynamic mixing with low spin (LS) impurities, and inter-molecular correlation, which is caused by elastic strain. 
Such a competition also leads to a nonzero macroscopic electric polarization, providing a mechanism for magnetoelectric coupling.

We study a microscopic model on a lattice in which each site represents one Mn(taa) molecule.
The $\mathrm{Mn}^{3+}$ cation as the center atom has two possible
spin-states: LS with $S=1$ and HS with $S=2$. 
The state of each site $i$ is labeled
by $\left(s_{i},m_{z,i},q_{i},\left\{ n_{k}\right\} _{i}\right)$,
where $s_{i}=l,h$ denotes the LS ($S=1$) and HS ($S=2$) state, and $m_{z,i}\in\left[-s_{i},s_{i}\right]$
is the magnetic quantum number.
$q_{i}=0$ for LS and $q_{i}=x,y,z$ for HS represent the elongation axis
of the JT distortion, and $\left\{ n_{k}\right\}_{i} $ is a set of vibration
numbers of the on-site phonon mode $k$ of Mn(taa). 
The indices $\left(s_{i},m_{z,i},q_{i}\right)$ denote the spin-JT states, with a total of eighteen states: 
three LS $S_{z} = 1,0,-1$ and fifteen HS ($S_{z} = 2,1,0,-1,-2$ and three JT elongation axes), for each site.
Therefore, the total Hamiltonian $\mathcal{H}$ is written in
three parts: the on-site term $\mathcal{H}_{\mathrm{ons}}$, the interaction
term $\mathcal{H}_{\mathrm{int}}$, and the electromagnetic term $\mathcal{H}_{\mathrm{EM}}$,
\begin{equation}
\mathcal{H}=\mathcal{H}_{\mathrm{ons}}+\mathcal{H}_{\mathrm{int}}+\mathcal{H}_{\mathrm{EM}}.
\end{equation}
The on-site term is,
\begin{equation}
\mathcal{H}_{\mathrm{ons}}=\sum_{i}\left[\Delta_{i}\left(s_{i}\right)+\sum_{k} 
\left(\hat{n}_{k}+{\textstyle\frac{1}{2}}\right)\text{\ensuremath{\hbar}}\omega_{k}\left(s_{i}\right)\right]
\end{equation}
where $\Delta(s_{i})$ is defined as the on-site energy difference between the LS and HS states, 
or the LS-HS gap; $\Delta$ is zero for $s_{i}=l$ and positive for $s_{i}=h$.
The next term is the on-site phonon energy. 

\begin{figure}
\includegraphics[width=1\columnwidth]{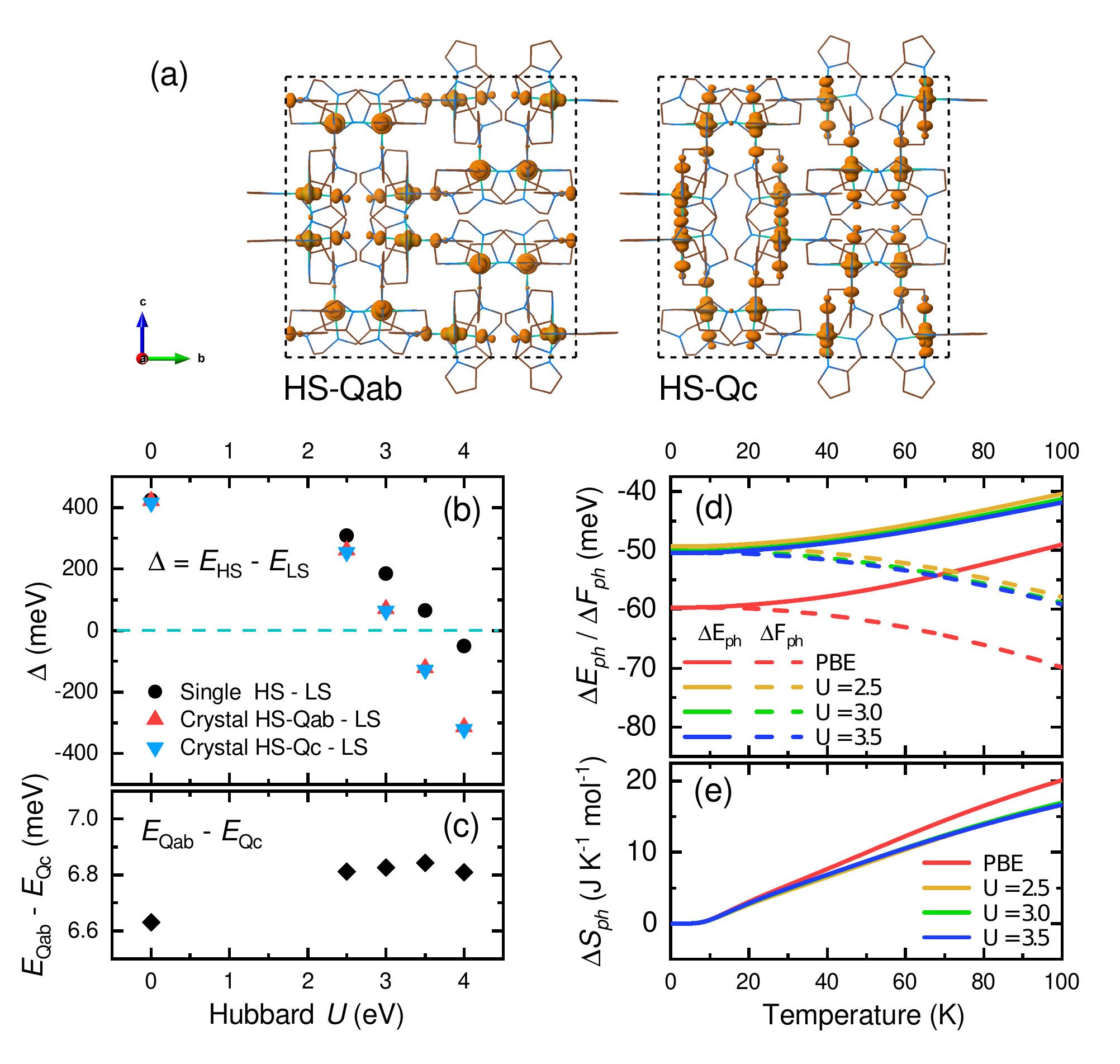}
\caption{(a) Isosurfaces of charge densities
of the highest sixteen occupied eigenstates in the spin-up channel of the 
two HS configurations HS-Q$_{ab}$ and HS-Q$_{c}$.
(b)(c) Energy difference per Mn atom 
between LS and HS states (b) as function of Hubbard $U$ for both
isolated molecule (single) and bulk (crystal) structure and 
(c) between HS-Q$_{ab}$ and HS-Q$_{c}$. 
(d) On-site phonon contributions to energy difference $\Delta E_{ph}$,
free energy difference $\Delta F_{ph}$, 
and (e) entropy difference $\Delta S_{ph}$ between LS and HS states 
as a function of the temperature.
}
\label{fig:dft}
\end{figure}

The LS-HS gap is derived directly from the total energy calculations
for LS and HS from first principles. 
The total energies are obtained for both the isolated molecules and
the bulk structures with LS and HS state using GGA-PBE and GGA+$U$,
varying $U$ from 2.5 eV to 4 eV. Two HS configurations in the bulk structure,
one labeled by HS-Q$_{ab}$ with $ab$-plane JT distortion and the
other labeled by HS-Q$_{c}$ with JT distortion along the $c$-axis, are
calculated. Fig.~\ref{fig:dft}(a)
shows the distribution of $3z^{2}-r^{2}$ centered molecular orbitals
of Mn(3d), which are the highest occupied orbitals due to the JT distortion.
The total energy difference between the LS and HS states per Mn atom as a function
of Hubbard $U$ depends strongly on $U$, as shown in Fig.~\ref{fig:dft}(b),
because the LS state, with an orbital occupied by two electrons, has an additional onsite energy $U$. In addition, the DFT ground state energy is higher for the LS state \cite{Chen_2015a}. 
For the purpose of this work, it is
necessary to tune $U$ to a value that can reproduce a reasonable
LS-HS gap, at about $ 100 \, \textrm{meV} $ \cite{Cirera_2018a}.
We find that $U\approx 3.0\sim3.5$ eV. As a reference,
we obtain from a constrained Random Phase Approximation (cRPA)\cite{Aryasetiawan_2004a,Aryasetiawan_2006a}
calculation using the FP-LAPW method\cite{Kozhevnikov_2010a,Zhang_2019a}
an estimated $U \approx 3.2 \, \textrm{eV} $ for Mn(3$d$) electrons
in Mn(taa) \cite{suppl}.

The frequencies of the 159 phonon modes, including three translation modes and three
rotation modes with zero frequency, are calculated for the isolated Mn(taa) in both LS and HS states. Certain phonon modes are softened in the HS state because
one of the Mn-N bonds is elongated due to JT distortion, and correspondingly their occupation
numbers are increased. 
Fig.~\ref{fig:dft}(d)(e) show the results of differences of energy, free
energy and entropy of phonon as a function of temperature by GGA-PBE
and GGA+$U$. At zero
temperature $T$, with zero vibration number for all phonon modes,
the zero-point energy difference $E_{ph}(h)-E_{ph}(l)$ is
$ -59.2 \, \mathrm{meV} $ for PBE and varies between $ -49.5\sim-59.7 \, \mathrm{meV} $ 
for GGA+$U$ with $U = 2.5\sim3.5 \, \, \mathrm{eV} $, respectively.
The phonon entropy is insensitive to $U$ and is larger for HS than LS, with
the difference increasing with $T$.

The total entropy difference including both spin and phonon contributions causes the SCO to become
a first-order phase transition.\cite{Sim_1981a,Garcia_2000a}
The transition temperature $T_{c}$ can be controlled by
external strain\cite{Guionneau_2005a} and magnetic field\cite{Garcia_2000a,Kimura_2005a,Her_2012}.
Although it was suggested that the phonon entropy does not dominate the phase transition on the basis of solid-state Raman spectra,\cite{Nakano_2003a}
a previous DFT study \cite{Garcia_2007a} found that the phonon entropy difference in Mn(taa) 
is comparable to the spin-JT entropy difference, which is 
$k_{B}\ln(5\times 3/3)=13.8 \, \mathrm{ J\cdot K^{-1}mol^{-1}}$ 
at $T_{c}$.
At $T_{c} = 45 \, K$, we calculate the phonon entropy difference to be $7.5\sim 8.8 \, \mathrm{J\cdot K^{-1}mol^{-1}}$.

Inter-molecular interactions arise from the elastic energy, 
which is treated classically because of the low vibration frequencies and high occupation numbers. 
Boukheddaden et al. 
\cite{Boukheddaden_2007a,Nishino_2007a,Slimani_2015a,Traiche_2018a}
suggested an isotropic elastic energy, which works well
for many 1D and 2D lattice SCO systems. In the case of Mn(taa), because
of the JT-active HS state, the strain interaction is
anisotropic and depends on the shape of the molecules.
Therefore, we assume an elastic energy of the form, 
\begin{equation}
\mathcal{H}_{\mathrm{int}}=\sum_{\left\langle i,j\right\rangle } A_{s_{i}s_{j}} \, [\left(\mathbf{Q}\left(q_{i}\right)+\mathbf{Q}\left(q_{j}\right)\right)\cdot\hat{\mathbf{r}}_{ij}]^{2}
\end{equation}
where $\mathbf{Q}$ is a dimensionless vector along the
JT elongation axis so that $\mathbf{Q}\left(l,0\right)=\left(0,0,0\right)$,
$\mathbf{Q}\left(h,x\right)=\left(1,0,0\right)$, $\mathbf{Q}\left(h,y\right)=\left(0,1,0\right)$,
and $\mathbf{Q}\left(h,z\right)=\left(0,0,1\right)$. Each Mn(taa) molecule has five nearest neighbors. 
$\hat{r}_{ij}$ is a unit vector between two neighbor sites $i$ and $j$.
$A$ is the force constant in the dimension of energy and is a function
of the spin-state $S$.  We assume that in the LS state, two neighbor molecules are separated by the equilibrium distance with zero elastic energy, thus $A_{ll}=0$. 
$A_{hh}$ is determined by the total energy difference
between two HS bulk calculations with HS-Q$_{ab}$ and HS-Q$_{c}$ as indicated in Fig.~\ref{fig:dft}(a). The
plot in Fig.~\ref{fig:dft}(c) shows that HS-Q$_{c}$ is always lower
in energy than HS-Q$_{ab}$ so that the JT elongation axes of HS tend
to align in parallel with each other. 
We find $A_{hh}=-10.0 \, \mathrm{meV}$, and to be insensitive to $U$. The negative sign indicates an
effective anisotropic attraction caused by the JT distortion between
two neighboring molecules. 

The effects of magnetic field $\mathbf{H}$, magnetic anisotropy $D$,
and external electric field $\mathbf{E}$ are given by $H_{\mathrm{EM}}$,
\begin{equation}
\mathcal{H}_{\mathrm{EM}}=-\sum_{i}[\mu_{0}\mathbf{H}\cdot\mathbf{m}_{i} + D \left(\mathbf{\hat{M}}\left(q_{i}\right)\cdot\mathbf{m}_{i} \right)^{2}+\mathbf{E}\cdot\mathbf{P}_{i}]
\end{equation}
where $\mathbf{m}_{i}$ is the local magnetic moment on each site, whose $z$ component is defined by the quantum number $m_{z,i}$, and $\mathbf{P}_{i}$ is the electric dipole. 
$\mathbf{\hat{M}}(q_i)$ is a unit vector along the JT elongation axis determined by the JT state $q_{i}$ to account for the excess anisotropy contribution  
in the HS state over the LS state. 
The magnetic anisotropic energy per molecule is $ 3.7~\, \mathrm{meV} $ from first-principles,  
yielding $D\approx 0.3 \, \mathrm{meV} $. In the Monte Carlo simulation,
we assume that the magnetic moments are always aligned with $\mathbf{H}$, which is in the $z$ direction. Thus the anisotropy energy only
needs to be computed for
sites with $q_{i} = z$.

\begin{figure}[ht]
\includegraphics[width=0.90\columnwidth]{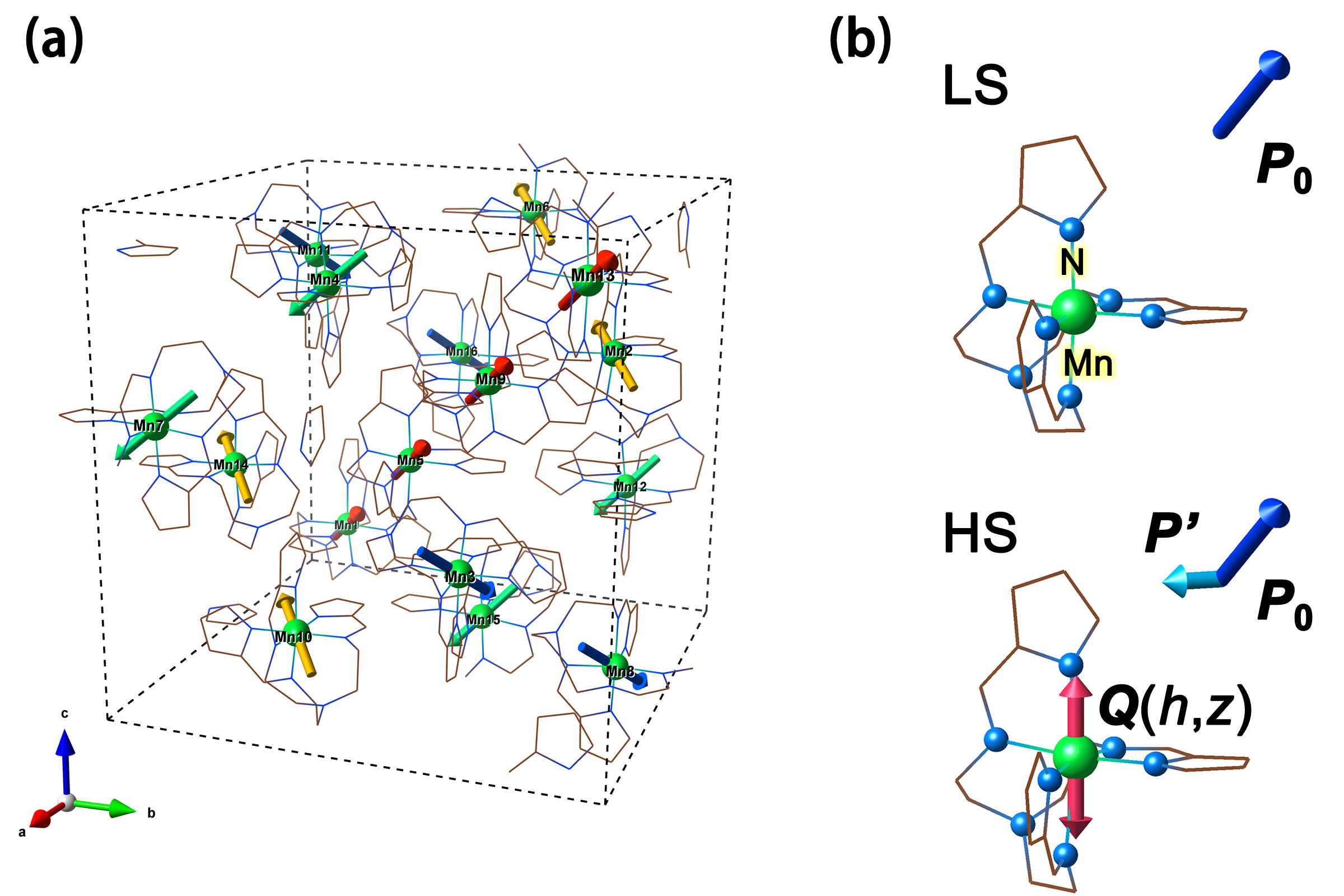}

\caption{Electric dipole of Mn(taa). (a)
LS state dipole directions (arrows) in a unit cell.
Four colors represent four groups of molecules with similar orientations. 
(b) For a single Mn(taa) molecule, the LS dipole $\mathbf{P}_{0}$
(blue arrows) is along the {[}111{]} direction and cancels out across the four groups. 
In the HS state, an additional $\mathbf{P}^{'}$ (light blue arrow) is perpendicular
to both the JT distortion $\mathbf{Q}(h,z)$ (red double-headed
arrow along the $z$ axis) and $\mathbf{P}_{0}$. }
\label{fig:polar}
\end{figure}

First principles calculations yield the magnitude $|\mathbf{P}|$ of about 1.15~$e$\AA~ for both LS and HS states. 
In the LS state, the dipole moment directions of the molecules in the unit cell are shown in Fig.~\ref{fig:polar}(a).
Sixteen molecules belong to four groups, 
with the dipole directions in each group aligned along one of the four $\left[111\right]$ directions. 
There is no net polarization in the unit cell. 
Once one molecule transitions to the HS state, an
additional transverse dipole $\mathbf{P}^{'}$ of about 0.23~$e$\AA~ is created along the
direction perpendicular to both its JT elongation axis and the original dipole
direction, as shown in Fig.~\ref{fig:polar}(b).

\begin{figure}[ht]
\includegraphics[width=0.95\columnwidth]{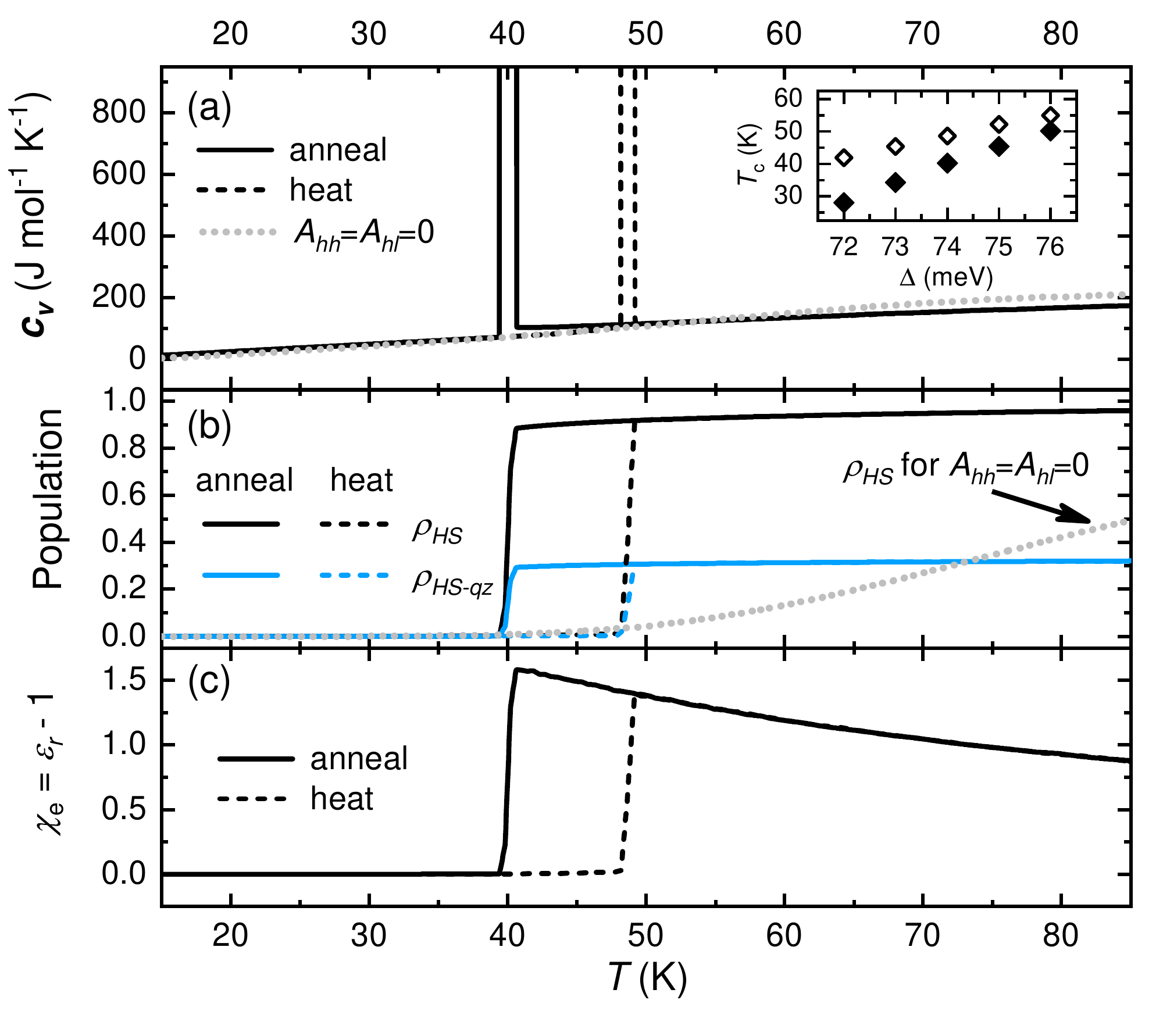}

\caption{Monte Carlo results under zero magnetic field for both cooling (solid) and heating (dashed). 
(a) specific heat $c_{v}$, (b) HS population $\rho_{\mathrm{HS}}$ and HS sub-population for $q=z$, $\rho_{\mathrm{HS}-qz}$, 
and (c) electric susceptibility $\chi_{e}$ as a function of temperature. 
The parameters used are $\Delta=74~\mathrm{meV}$, $A_{hh}=-5.0~\mathrm{meV}$,
$A_{hl}=5.0~\mathrm{meV}$ and $D=0.3~\mathrm{meV}$. 
Gray dotted curves in (a) and (b) are results with the same parameters but with zero strain interaction ($A_{hh}=A_{hl}=0$).}
Inset in (a) gives the SCO transition temperature as a function of $\Delta$ for cooling (solid) and heating (hollow).
\label{fig:mc_zero}
\end{figure}

The first-order SCO phase transition as a function of $T$ in $\mu_{0}H = 0$ is reproduced by
Monte Carlo simulation using the microscopic Hamiltonian with the parameters obtained from first-principles calculations. 
The parameters are $\Delta=74\mathrm{~meV}$, $A_{hh}=-5.0~\mathrm{meV}$, $A_{hl}=5.0~\mathrm{meV}$, and $D=0.3~\mathrm{meV}$. 
A small constant electric field of $0.1$ mV/\AA~ is applied along the $z$ direction. 
In Fig.~\ref{fig:mc_zero} we show $\mu_{0}H=0$ specific heat $c_v$ (a),
HS population $\rho_{\mathrm{HS}}$ and sub-population for $q=z$,
$\rho_{\mathrm{HS}-qz}$ (b), and
electric susceptibility $\chi_{e}$ (c) as a function of temperature. The first-order transition is
clearly indicated by the
singularity in the specific heat $c_{v}$ and the discontinuity in $\rho_{\mathrm{HS}}$,
$\rho_{\mathrm{HS}-qz}$, and $\chi_e$ at the same $T$.
$T_{c}$ averaged over cooling and heating is 44 K, 
matching experiments.\cite{Sim_1981a,Garcia_2000a} 
Without the elastic energy, the LS to HS transition occurs gradually with increasing temperature, and no phase boundary can be observed (gray dotted curves in Fig.~\ref{fig:mc_zero}(a)(b)). 
This confirms the key role of the inter-molecular strain interaction in the SCO phase transition. \cite{suppl}

\begin{figure}[ht]
 \includegraphics[width=1\columnwidth]{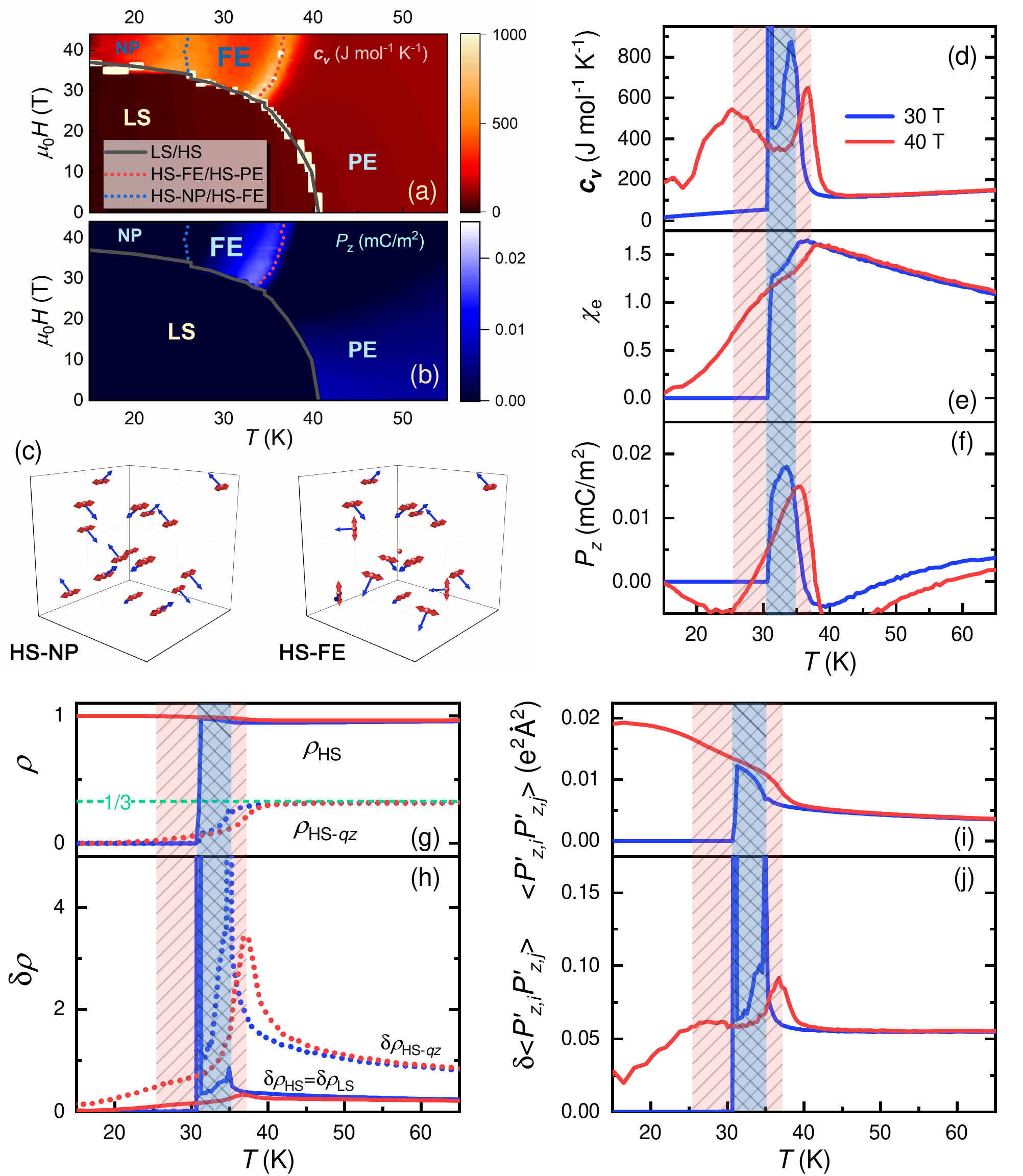}

 \caption{Monte Carlo results in finite $\mu_{0}H$. 
Phase diagrams obtained from (a) specific heat $c_{v}$ and 
(b) $z$-component of polarization $P_{z}$, 
as a function of $T$ and $\mu_{0}H$. Phases are labeled by
LS, NP (HS non-polarized), FE (HS ferroelectric), and PE (HS paraelectric), respectively. 
(c) For NP and FE, snapshots of JT-elongation axis $\mathbf{Q}$ (red double-headed arrows) and the corresponding transverse dipoles $\mathbf{P}^{'}$ 
(blue arrows) in a unit cell with sixteen sites. 
In $ 30 \T $ (blue) and $ 40 \T $ (red) fields, 
(d) $c_{v}$, 
(e) $\chi_{e}$, (f) $P_{z}$, 
and the thermal average and the corresponding fluctuation (standard deviation) of (g)(h) $\rho_{\mathrm{HS}}$ and $\rho_{\mathrm{HS}-qz}$ 
and (i)(j) nearest neighbor correlation $\langle P^{'}_{z,i}P^{'}_{z,j}\rangle$  
as a function of $T$.  
The shaded regions mark the FE phase in $\mu_{0}H = 30$~T (blue) and $40$~T (red).
}
\label{fig:mc_phase}
\end{figure}

Now we are ready to examine the three HS phases by performing the simulations under different magnetic fields.
The results are shown in Fig.~\ref{fig:mc_phase}.
All four phases, LS, NP (HS non-polarized), FE (HS ferroelectric), and PE (HS paraelectric), are identified with phase boundaries (a) defined by the peaks in $c_{v}$, discontinuities in
$\rho_{\mathrm{HS}-qz}$ and the $z$ component of the electric polarization $P_z$.
The LS phase is located at the low magnetic field and low temperature part of the phase diagram.
The boundary between the LS and HS phases is the first order SCO phase transition and the transition temperature approaches zero when the magnetic field reaches about 35~T.
The PE phase is the high temperature HS phase with diminishing electric polarization. The NP and FE phases only exist under high magnetic fields.

The low-temperature HS phase is non-polarized when temperature is below about $ 25 \, \textrm{K} $. 
The snapshot in Fig.~\ref{fig:mc_phase}(d) for this phase shows that all sites have the same JT-elongation axes so that the transverse dipoles $\mathbf{P}^{'}$ are exactly cancelled, 
leaving zero net polarization. 
The electric susceptibility $\chi_{e}$ decays quickly with decreasing temperature, 
indicating that the JT degrees of freedom are being frozen out. This is analogous to the freezing of a material into a crystalline solid. Thus we identify this phase as the JT solid phase.
The phase transition temperature of $ 25 \, \textrm{K} $ is determined by $A_{hh}$, 
the strength of the HS-HS strain interaction \cite{suppl}.

The high-temperature HS phase, labeled PE,
shows an almost ideal sub-population of all three JT axes, each at close to $1/3$. 
Because the three JT axes are chosen at random with almost no correlation, 
the electric dipole and susceptibility arise from thermal fluctuation.
This leads to a paraelectric behavior. 
The nearly uncorrelated JT distortion is analogous to a gas. 
Thus we identify the PE phase as a JT gas phase.

The intermediate-temperature HS phase, labeled FE, has a non-zero net polarization. 
The snapshot of the FE phase unit cell in Fig.~\ref{fig:mc_phase}(d) shows one $q=0$, three $q=x$, nine $q=y$ and three $q=z$ molecules.
Recall that $q=0$ is the LS state. This means that
the FE phase contains a small number of LS molecules.
The sub-population of each JT axis is below its nominal value $1/3$, also plotted in Fig.~\ref{fig:mc_phase}(f).
This mixing of a small number of LS molecules naturally introduces disorder in JT ordering.
Consequently the dipoles are no longer compensated,
The electric susceptibility increases with increasing temperature and has a maximum at the transition temperature. 
This is a typical feature of ferroelectrics below 
the Curie temperature \cite{Samara_1966}.
The thermal average polarization in this phase is about 
$ 0.02 \, \mathrm{mC/m^{2}}$, smaller than that of one 
uncompensated dipole at a 10000 site. 
This strongly suggests that the nonzero polarization arises from dynamic fluctuation of the JT distortion, 
and that this intermediate phase is analogous to a liquid phase.
Thus, we identify it as a JT liquid.

To understand the nature of the three JT phases, and to confirm that the intermediate phase indeed has liquid characteristics, 
we calculate the thermal average and the fluctuation of the nearest neighbor polarization correlation function 
$\langle P^{'}_{z,i} P^{'}_{z,j}\rangle$, which are plotted in Figs.~\ref{fig:mc_phase}(i) and (j). 
The LS state without the JT degree of freedom has zero correlation and zero fluctuation in the low $T$ region in $\mu_{0}H = 30$~T.
Among the three HS phases, the thermal average of the correlation for the NP and FE phases is much larger than for the PE phase,
reflecting stronger correlation in the two lower temperature HS phases.
On the other hand, fluctuation is higher for the two higher temperature
HS phases, FE and PE, than the NP phase, besides the maximum at the phase transition boundary. 
The fluctuation in the FE phase is $\sim0.06 \,e^{2}$\AA$^{2}$ away from the phase boundary, much larger than the thermal average as $\sim0.01 \,e^{2}$\AA$^{2}$. 
In contrast, the fluctuation in the NP phase decays as $T$ decreases due to the frozen JT degree of freedom. 
The dynamic behaviors can also be confirmed by the fluctuation in population [Fig.~\ref{fig:mc_phase}(h)]; 
$\delta\rho_{\mathrm{HS}}$ is equal to $\delta\rho_{\mathrm{LS}}$.
The fluctuation $\delta\rho_{\mathrm{LS}}$ 
in the FE phase is larger than that in both LS and NP phases, indicating importance of the dynamics of the LS in the FE phase.
$\delta\rho_{\mathrm{HS-qz}}$ in the FE phase is also much larger than the corresponding thermal average.
The above analysis reveals that the FE phase exhibits correlated and dynamic JT distortion. 
Correlation and dynamic fluctuation are properties generally identified with a liquid in condensed matter.


In summary, we introduced a molecular-scale Hamiltonian
to describe the SCO system Mn(taa) with parameters obtained from first principles calculations. 
Monte Carlo simulations using this Hamiltonian qualitatively reproduce the experimental phase diagram and yield microscopic understanding
of the observed phases. 
The strain interaction is the primary factor controlling the ordering and dynamics of JT
distortions, producing three JT states of
matter (JT solid, JT liquid, and JT gas)
in different magnetic fields and temperatures.
The JT liquid is particularly interesting because
of the dynamic JT fluctuation in this phase
and the uncompensated electric polarization.
Disorder and dynamic JT fluctuation are quite
common in spin-crossover materials. Thus one
may expect JT liquid to be discovered in other
systems.

This work was supported as part of the Center for Molecular Magnetic Quantum Materials, 
an Energy Frontier Research Center funded by the U.S. Department of Energy, Office of Science, Basic Energy Sciences under Award No. DE-SC0019330. 
Computations were done using the utilities of National Energy Research Scientific Computing Center, 
the Extreme Science and Engineering Discovery Environment under No. TG-PHY170023
and University of Florida Research Computing systems.


\end{document}